\documentclass[twocolumn,aps,prb,floats,showpacs]{revtex4}
\usepackage{graphicx}
\usepackage{amsmath}

\newcommand{\rr}{{\bf r}}
\begin{document}
\title{Generalized Inverse Participation Ratio as a Possible Measure of Localization for Interacting Systems}
\author{N. C. Murphy}
\author{R. Wortis}
\author{W. A. Atkinson} \email{billatkinson@trentu.ca}
\affiliation{Department of Physics and Astronomy, Trent University, 1600 West Bank Dr., Peterborough ON, K9J 7B8, Canada}
\date{\today}
\begin{abstract}
We test the usefulness of a generalized inverse participation ratio
(GIPR) as a measure of Anderson localization.  The GIPR differs from
the usual inverse participation ratio in that it is constructed from
the local density of states rather than the single-electron
wavefunctions.  This makes it suitable for application to many-body
systems.  We benchmark the GIPR by performing a finite-size scaling
analysis of a disordered, noninteracting, three-dimensional
tight-binding lattice.  We find values for the critical disorder and
critical exponents that are in agreement with published values.
\end{abstract}
\maketitle

\section{Introduction}
Anderson localization is a phenomenon in which quantum particles may
be localized due to a random potential, even though the particles are
classically unbound.\cite{Anderson1958} The theory for noninteracting
particles is well-developed: in one and two dimensions, particles are
localized by arbitrarily weak disorder, and in three dimensions states
may be localized or extended depending on the strength of
disorder.\cite{Lee1985}

Most real particles are interacting, however, and there has been an
ongoing effort to understand how interactions modify the
noninteracting picture, either because of screening of the disorder
potential, or because of loss of quantum coherence due to inelastic
scattering.\cite{Altshuler1985} Until recently, neither of these
effects was believed sufficient to change the noninteracting picture
at zero temperature. However,
experiments\cite{Abrahams2001} in two dimensional semiconductor films
identified a zero-temperature metal-insulator transition (MIT) that
appears to result from electron interactions.\cite{Punnoose2005} More
recently, it has been suggested that weakly-interacting one and
two-dimensional systems will exhibit a finite-$T$ Anderson MIT.\cite{Gornyi2005,Basko2006}

There is also interest in Anderson localization in strongly
interacting systems.\cite{Miranda2005} Many of the most interesting
strongly correlated materials are insulators, but can have their
electronic properties tuned by chemical doping.  Of particular
interest are materials, such as the high temperature superconductors,
whose parent compounds have an interaction-driven Mott insulating
phase.  These materials become superconductors when doped with a few
percent of electron or hole donor atoms, but pass through various
intermediate phases in which disorder seems to play an important role.
There is an abundancy of questions about how the electronic
properties of these materials are modified by doping-related disorder.
Of particular relevance to this work, there have been recent questions
about how localization physics is altered near the Mott
MIT,\cite{Tanaskovic2003,Heidarian2003,Chakraborty2007,Henseler2008,Song2008,Henseler2009}
and about the phase transition between the Anderson and Mott
insulating phases.\cite{Byczuk2005,Semmler2010,Shinaoka2009b}

Finally, trapped atomic gases in random optical lattices have now been
experimentally realized.\cite{Chen2008,Fallani2007,White2009,Deissler2010,Modugno2010}  These systems
are interesting because the strength of the atom-atom interactions can
by tuned by application of an external magnetic field.  There is
therefore the prospect of making a controlled study of Anderson
localization as a function of interaction strength.

Numerical calculations have played an important role in understanding
Anderson localization in noninteracting systems. However, many of the
techniques developed for measuring localization in noninteracting
systems cannot be extended to interacting systems since they require
knowledge of the {\em single-particle} eigenstates of the system and,
with the exception of self-consistent field calculations, many-body
wavefunctions cannot generally be written as a simple product of
single-particle states.  There is, therefore, an interest in
developing new numerical methods for studying the Anderson MIT in
interacting systems.

With this in mind, there have been several proposals that the
localization transition can be detected by studying the statistical
properties of the local density of states (LDOS) $\rho({\bf
  r},\omega)$.  The geometric average of the LDOS, $\rho_g(\omega)$,
is an order parameter for the Anderson MIT in the limit of infinite
system size\cite{Janssen1998,Dobrosavljevic2003} because it vanishes
when the local spectrum is discrete.  In infinite systems, this occurs
only at energies at which the states are localized and not at which the
states are extended.
A generalization of dynamical mean field theory based on incorporating
$\rho_g(\omega)$ into the self-consistency cycle was developed to
study interacting disordered
systems.\cite{Dobrosavljevic2003,Byczuk2005,Semmler2010} As a
practical measure of localization in finite systems, however,
$\rho_g(\omega)$ is problematic because the spectrum is always
discrete, and this can obscure the Anderson
MIT.\cite{Song2007,Wortis2008}
More recently, several groups have suggested that the Anderson
transition can be detected by studying the distribution of $\rho({\bf
  r},\omega)$
values,\cite{Tran2007,Semmler2010,Schubert2010,Rodriguez2010} and it
has been shown that this distribution scales differently with system
size for localized and delocalized states.\cite{Schubert2010}

In this work, we consider a quantity, the generalized
inverse participation ratio (GIPR), that is related to the LDOS via
\begin{equation}
G_2(\omega) = \frac{\sum_i \rho({\bf r}_i,\omega)^2}
{[\sum_i \rho({\bf r}_i,\omega)]^2}.
\label{eq:GIPR}
\end{equation}
Equation~(\ref{eq:GIPR}) is defined for a lattice, so that $\rho({\bf
  r}_i,\omega)$ is the density of states projected onto the local
Wannier orbital at the $i$th site of the lattice.  The GIPR was used
previously in finite size scaling studies,\cite{Song2008} but a
careful examination of its scaling properties has not been made.
This is the purpose of this paper.

The GIPR is analogous to the usual inverse
participation ratio (IPR) for noninteracting systems,
\begin{equation}
I_{q,\alpha} = \frac{\sum_i|\Psi_\alpha({\bf r}_i)|^{2q}} {[\sum_i
    |\Psi_\alpha({\bf r}_i)|^2]^q},
\label{eq:IPR}
\end{equation}
where $\Psi_\alpha(\rr_i)$ is a single-particle wavefunction with quantum
number $\alpha$ in the
basis of Wannier orbitals.  The IPR is conventionally defined with
$q=2$ and can be used to distinguish Anderson localized and extended states:
for a finite $d$-dimensional system of linear size $L$,
$I_{2,\alpha}$ satisfies
\begin{equation}
\lim_{L\rightarrow \infty} I_{2,\alpha} = \left \{ \begin{array}{ll}
1/L^d &\mbox{(extended states)} \\
\mbox{const.} & \mbox{(localized states)},
\label{eq:IPRscaling0}
\end{array} \right .
\end{equation}
for states that are far from the Anderson MIT, and exhibits
multifractal scaling,\cite{Janssen1998,Mirlin2000,Mildenberger2002,Brndiar2006}
\begin{equation}
\lim_{L\rightarrow \infty} I_{2,\alpha} = L^{-d_2} \tilde F[(W-W_c)L^{1/\nu}],
\label{eq:IPRscaling}
\end{equation}
near the transition. Here, $d_2$ is the fractal dimension for $q=2$, $\nu$ is a
critical exponent, and $W$ and $W_c$ are the disorder and critical
disorder strengths respectively.

For noninteracting systems,  $G_2(\omega)$ reduces to the IPR when $\omega$
is equal to one of the eigenenergies of the system.
This follows from substituting
\begin{equation}
\rho({\bf r}_i,\omega) = \sum_\alpha |\Psi_\alpha({\bf
  r}_i)|^2\delta(\omega-E_\alpha),
\label{eq:LDOS}
\end{equation}
into Eq.~(\ref{eq:GIPR}), where $E_\alpha$ are the discrete
eigenenergies of the disordered lattice. However, for a general value
of $\omega$ not equal to one of the eigenenergies, $G_2(\omega)$ is
not well defined if the $\delta$-functions in Eq.~(\ref{eq:LDOS}) are
infinitely sharp, and the relationship between the IPR and the GIPR is
therefore ambiguous.  Moreover, we show below that if one broadens the
$\delta$-functions by an amount $\gamma$, there is no limiting value
of $\gamma$ in which the GIPR reduces to the IPR.  The goal of this
paper is to demonstrate that the GIPR can nonetheless be used to detect the
Anderson MIT and to determine the critical parameters $W_c$, $d_2$ and
$\nu$.

We benchmark the GIPR by performing finite size scaling for a
disordered noninteracting model, where the critical properties are
well known.  In Sec.~\ref{sec:calculations}, we discuss how the
broadening of the $\delta$-functions in Eq.~(\ref{eq:LDOS}) is expected
to affect the finite size scaling, and use this to select an optimal
broadening.  In Sec.~\ref{sec:results}, we show the results of
numerical finite size scaling, from which we extract values for
the critical disorder and critical exponents at the Anderson MIT.
We show that, with an appropriate choice for $\gamma$, it is possible to 
extract critical properties.

\section{Calculations}
\label{sec:calculations}

The noninteracting Anderson model is
\begin{equation}
\hat H = -t\sum_{\langle i,j\rangle} |i\rangle\langle j| +\sum_i 
|i\rangle \epsilon_i  \langle i |.
\end{equation}
where $|i\rangle$ is the ket for a Wannier orbital at position $i$ on
the lattice and $\langle i,j\rangle$ indicates that the sum is over
nearest-neighbour sites. The hopping matrix element is taken to be
$t=1$, and it therefore sets the energy scale, while the site energies
$\epsilon_i$ are taken from a uniform distribution of random values
ranging from $-W/2$ to $W/2$, where $W$ is the strength of disorder.
Calculations are performed for a three-dimensional ($d=3$) cubic
lattice of linear size $L$ and with $N_s=L^3$ lattice points.

We use a recursion method\cite{Haydock1994} to find the local Green's
function $G({\bf r}_i,\omega+i\gamma)$ at site $i$, where $\gamma$ is
a small but finite shift off the real frequency axis.  
This method introduces an error through truncation of the recursion
algorithm, and we have been careful to adjust the truncation criterion so
that this error is much smaller than the error due to disorder averaging.
The LDOS is
given by the imaginary part of $G({\bf r}_i,\omega+i\gamma)$.
Formally, this is equivalent to
\begin{equation}
\rho_\gamma({\bf r}_i,\omega) = \frac{1}{\pi}\sum_\alpha |\Psi_\alpha({\bf
  r}_i)|^2\frac{\gamma}{(\omega-E_\alpha)^2+\gamma^2}
\label{eq:LDOS2}
\end{equation}
where $E_\alpha$ are the eigenenergies for a particular disorder realisation.
Once $\rho_\gamma(\rr_i,\omega)$ is
known, the GIPR is calculated from Eq.~(\ref{eq:GIPR}).  In this work,
we focus on the band center ($\omega=0$), where the Anderson transition
is well-characterized.  In particular, the Anderson MIT occurs at
a critical disorder $W_c = 16.5t$ for the uniform disorder distribution used
here.\cite{MacKinnon1981,Slevin1999}

\begin{figure}
\includegraphics[width=\columnwidth]{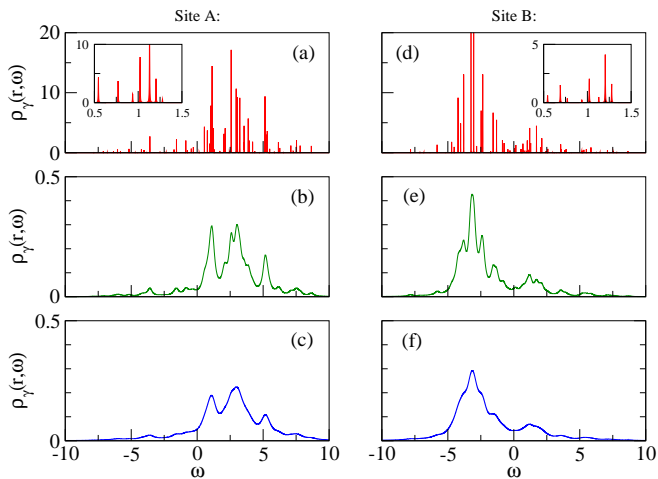}
\caption{(Color online) Local density of states for two well-separated
  lattice sites, ``A'' and ``B'', in a disordered lattice.  Panels
  show the LDOS at (a)-(c) A and (d)-(f) B.  All spectra are for the
  same configuration of disorder, but have different values of
  $\gamma$.  Results are for (a), (d) $\gamma=0.01W/N_s$; (b), (e)
  $\gamma=W/N_s$; (c), (f) $\gamma=2W/N_s$, where the lattice has $N_s
  = 4^3$ sites and $W=13$.  Insets show expanded views of the LDOS
  near $\omega=1$.}
\label{fig:ds_dG}
\end{figure}

One of the main issues we face is how to choose $\gamma$. 
In the remainder of this section, we
discuss how this choice affects both the LDOS and the scaling behavior of the
GIPR.  The relevant energy scale for comparison
is the level spacing at the band center, $\Delta = 1/\rho_0N_s$,
where $\rho_0$ is the system-averaged density of states at $\omega=0$.
For strongly disordered systems, $\Delta \approx W/N_s$,
while for weakly disordered systems, $\Delta \approx D/N_s$, where
$D$ is the bandwidth of the disorder-free lattice.  For the cubic lattice
considered here, the Anderson transition occurs at an intermediate
disorder strength, so that $\Delta$ lies between these two limits.
%

Figure \ref{fig:ds_dG} shows the dependence of the LDOS on
$\gamma$. When $\gamma \gtrsim \Delta$, the LDOS at $\omega$ is an
average over states with $|\omega-E_\alpha|\lesssim \gamma$, with the
consequence that the LDOS is more spatially uniform than the
individual eigenstates making up the LDOS.  Thus in
Fig.~\ref{fig:ds_dG}, the sites A and B are spatially separated, and
both have broad peaks at $\omega = 1$.  It is not possible to tell,
based on the LDOS for $\gamma =W/N_s$, whether these peaks indicate a
single eigenstate or a cluster of eigenstates that happen to be close
in energy.  It is only when $\gamma \ll \Delta$ that we see that the
local spectrum is quite different at the two sites
[Figs.~\ref{fig:ds_dG}(a) and (d)].
This suggests that the 
finite size scaling of the 
LDOS, and by extension the GIPR, should do a better
job of distinguishing localized and extended states as $\gamma$ is
reduced.

However, the fact that the spectrum develops discrete peaks when
$\gamma \ll \Delta$ does not mean that the LDOS samples only
individual eigenstates in this limit. This is because most energies do
not coincide with a peak.  When $\gamma \ll \Delta$, the energy
$\omega=0$ lies in the tails of the surrounding peaks and
Eq.~(\ref{eq:LDOS2}) becomes
\begin{equation}
\rho_\gamma(\rr_i,0) = \frac{\gamma}{\pi} \sum_{\alpha}
\frac{|\Psi_{\alpha}(\rr_i)|^2}{E_\alpha^2}.
\label{eq:rhosmallgamma}
\end{equation}
This means that even in the limit $\gamma\rightarrow 0$, $\rho(\rr_i,0)$
 is averaged over a nonzero number of states.  The LDOS at
$\omega=0$, and by extension the GIPR, does not change qualitatively when 
$\gamma$ is reduced much below $\Delta$.

We can learn more about the GIPR scaling by 
substituting Eq.~(\ref{eq:LDOS2}) into Eq.~(\ref{eq:GIPR}), from which
we obtain
\begin{eqnarray}
G_2(0) &=& \sum_i \left (\sum_\alpha w_\alpha |\Psi_\alpha(\rr_i)|^2 \right )^2 
\label{eq:GIPR1} 
\end{eqnarray}
where
\begin{equation}
w_\alpha = \frac{(E_\alpha^2 + \gamma^2)^{-1}}{\sum_\beta(E_\beta^2 + \gamma^2)^{-1}}
\end{equation}
is a weighting factor satsifying $\sum_\alpha w_\alpha = 1$.  
In the limit of vanishing disorder, the wavefunctions are plane waves
with $|\Psi_\alpha(\rr_i)|^2 = N_s^{-1}$, and Eq.~(\ref{eq:GIPR1})
gives $G_2(0) = N_s^{-1}$; this result is independent of $\gamma$ and
is identical to the scaling result for the IPR.

In the limit of large disorder, $W\gg W_c$, it is useful to rearrange
Eq.~(\ref{eq:GIPR1}) to obtain,
\begin{eqnarray}
G_2(0) &=& \sum_\alpha w_\alpha^2 I_{2,\alpha}
+ \sum_{\alpha\neq\beta} w_\alpha w_\beta \sum_i |\Psi_\alpha(\rr_i)|^2|\Psi_\beta(\rr_i)|^2.
\nonumber \\
\label{eq:GIPR2}
\end{eqnarray}
The first term on the right hand side is a weighted sum of IPR values
for eigenstates with $|E_\alpha|\lesssim \gamma$, while the second
term consists of cross terms between pairs of eigenstates.  The second
term can be neglected when the distances between these localized states are
large compared to the localization length $\xi$.  We can estimate
the typical distance
between centers of localization of the states in Eq.~(\ref{eq:GIPR2})
for the case $\gamma\gtrsim\Delta$.  In this case, there 
are of order $2\gamma/\Delta$ states with 
 $|E_\alpha|<\gamma$, and the mean separation of these is
\begin{equation}
\ell \sim L(\Delta/2\gamma)^{1/d}.
\label{eq:ell}
\end{equation}
The product
$|\Psi_\alpha(\rr_i)|^2|\Psi_\beta(\rr_i)|^2$ for two states separated
by $\ell$ has a maximal value of order
  $\exp[-2L(\Delta/2\gamma )^{1/d}/\xi]$, at the midpoint between
the centers of localization.
It follows that the second term in Eq.~(\ref{eq:GIPR2}) vanishes for
$L/\xi\rightarrow \infty$, in which limit the GIPR is expected to
scale like the IPR.  

For finite $L$, however, the second term in
Eq.~(\ref{eq:GIPR2}) introduces finite size corrections to the GIPR
that make it scale differently from the conventional IPR.  In order to
minimize these corrections, we want to make $\ell$ as large as
possible, which is achieved by taking $\gamma$ as small as possible.
We emphasize, however, that Eq.~(\ref{eq:ell}) only holds for $\gamma
\gtrsim \Delta$, and that $w_\alpha$
is independent of $\gamma$ when $\gamma \ll \Delta$,
namely
\begin{equation}
\lim_{\gamma\rightarrow 0} w_\alpha = \frac{E_\alpha^{-2}}{\sum_{\beta} E_{\beta}^{-2}}.
\end{equation}
In other words, $\ell$ ceases to increase when $\gamma$ is much less
than $\Delta$.  Our analysis therefore suggests that one cannot do
much better at minimizing finite size effects than by taking $\gamma
\sim \Delta$.

Finally, having established that $G_2(0)$ is determined by the first term in
Eq.~(\ref{eq:GIPR2}) when $L/\xi \gg 1$, we show that the weighting
terms do not affect the GIPR scaling in this limit.  We write
$I_{2,\alpha} \approx I_2(E_\alpha)$, where $I_2(E)$ is a slowly varying
function of $E$ near $E=0$, so that 
\begin{equation}
G_2(0) \approx I_{2}(0) \sum_\alpha w_\alpha^2.
\label{eq:G0scaling}
\end{equation}
For $\gamma \gtrsim \Delta$, we may estimate the sum 
over eigenstates by 
\begin{equation}
\sum_\alpha w_\alpha^2 \approx \frac{\Delta^{-1}\int dE
  (E^2+\gamma^2)^{-2}}{\left [ \Delta^{-1} \int dE (E^2 +
    \gamma^2)^{-1} \right ]^2} \propto \frac{\Delta}{\gamma}.
\label{eq:wa2}
\end{equation}
Taking $\gamma \propto N_s^{-1}$ eliminates the $L$-dependence of
the weighting factors in Eq.~(\ref{eq:GIPR2}).

In summary, we have shown that the GIPR will reproduce the scaling of
the IPR in the limits of vanishingly weak and strong disorder.
Moreover, we have shown that using smaller $\gamma$ values to
calculate the GIPR is preferable, down to $\gamma\sim \Delta$.  Many
numerical methods converge faster for larger $\gamma$ and, for these,
$\gamma \sim \Delta$ will be optimal.  In the next section, we examine
whether the finite size effects near $W_c$ limit our ability to
extract the critical behaviour.

\section{Results}
\label{sec:results}

\begin{figure}
\includegraphics[width=\columnwidth]{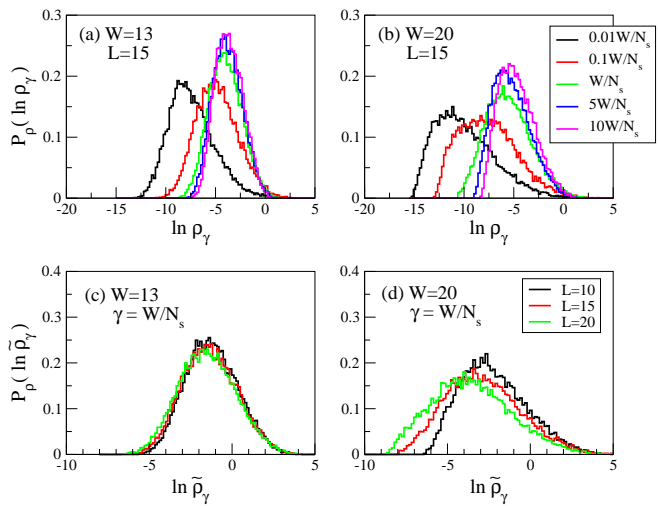}
\caption{(Color online) Probability distribution of the logarithm of
  the local density of states at $\omega=0$.  The effect of $\gamma$
  on $P_\rho(\ln \rho_\gamma)$ is shown for (a) extended and (b)
  localized states for a fixed system size.  The effect of system size
  on $P_\rho(\ln \tilde \rho_\gamma)$ is shown for (c) extended and
  (d) localized states for $\gamma = W/N_s$.  Here, $\tilde
  \rho_\gamma$ is the normalized LDOS, $\tilde \rho_\gamma \equiv
  \rho_\gamma/\langle\rho_\gamma\rangle$, where
  $\langle\rho_\gamma\rangle$ is the system-averaged LDOS at
  $\omega=0$.  Results are shown for 16 ($L=10$), 5 ($L=15$), and 2
  ($L=20$) disorder configurations, such that the number of LDOS values
  in each case is roughly the same.}
\label{fig:lnLDOS}
\end{figure}

We plot, in Fig.~\ref{fig:lnLDOS}, the probability distribution of the
logarithm of the LDOS at $\omega=0$ for different values of $\gamma$
and for different system sizes.  Figures~\ref{fig:lnLDOS}(a) and (b)
show that $\gamma$ affects both the peak position and shape of the
distribution.  In particular, the peak position of the distribution
$P_\rho(\ln \rho_\gamma)$ is proportional to $\gamma$ for $\gamma \lesssim
W/N_s$, in accordance with Eq.~(\ref{eq:rhosmallgamma}).  For
$\gamma\gtrsim W/N_s$, the peak position and width are weak functions
of $\gamma$.  

In Figs.~\ref{fig:lnLDOS}(c) and (d), we show the $L$-dependence of
the distribution of the normalized LDOS, $\tilde \rho_\gamma \equiv
\rho_\gamma/\langle \rho_\gamma\rangle$ with $\langle
\rho_\gamma\rangle$ the sample-averaged density of states.  Schubert
et al.\cite{Schubert2010} showed that the scaling of the distribution
$P_\rho(\ln \tilde \rho_\gamma)$ can be used to distinguish localized
and extended states: $P_\rho(\ln \tilde \rho_\gamma)$ shifts to the
left with increasing $L$ for localized states, and is independent of
$L$ for extended states.  Here, we find that there is indeed a
pronounced shift for the localized case ($W=20$), and that the
distribution is almost independent of $L$ for the extended case
($W=13$). The small leftward shift seen in the extended case is
presumably due to finite-size effects, which are more pronounced here
than in Ref.~\onlinecite{Schubert2010}.  Despite its smallness, this
leftward shift is problematic because it obscures the signature of the
Anderson transition in $P_\rho(\ln \rho_\gamma)$.  This is a
potentially important issue for many-body calculations where
accessible system sizes tend to be severely limited.  It appears that,
as with other measures of localization, the usefulness of the LDOS
distribution will depend on the inclusion of finite-size corrections.

\begin{figure}
\includegraphics[width=\columnwidth]{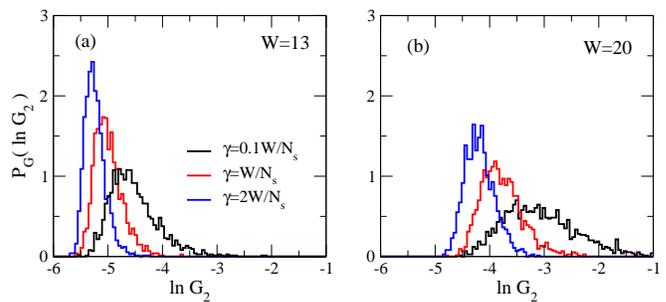}
\caption{(Color online) Effect of $\gamma$ on the probability
  distribution of the GIPR at $\omega=0$.  Results are for (a)
  extended ($W=13$) and (b) localized ($W=20$) states, and are for
  1500 disorder configurations with $L=10$.}
\label{fig:change_gamma}
\end{figure} 

Figure \ref{fig:change_gamma} shows the probability distribution
function $P_G(\ln G_2 )$ for the GIPR, obtained by calculating
$G_2(\omega)$ at $\omega=0$ for 1500 distinct impurity configurations.
This figure shows that the width of the distribution depends strongly
on $\gamma$, and that $P_G(\ln G_2)$ is sharply peaked when $\gamma
\gtrsim \Delta$.  Because the distribution of $\ln G_2$ is narrow, the
mean and most probable values of the distribution are close to each
other.  For this reason we study the finite size
scaling of the {\em typical} GIPR\cite{Mirlin2000b}
\begin{equation}
G_2^\mathrm{typ}(\omega) = \exp\left [ \langle \ln G_2(\omega)
\rangle \right ],
\end{equation}
 where $\langle \ldots \rangle$ refers to an
average over disorder configurations.  

We argued in the previous section that one should take $\gamma \propto \Delta$,
where $\Delta$ depends on both $N_s$ and $W$.  To understand whether the
$W$-dependence of $\Delta$ is important, we take two cases:
$\gamma \propto W/N_s$ and $\gamma \propto W_c/N_s$, where $W_c$ here refers
to the accepted value of 16.5.  As we discussed in Sec.~\ref{sec:calculations}, the first choice overestimates the $W$-dependence of $\Delta$, while
the second choice underestimates it.  Note that there is nothing
fundamental about the proportionality constant $W_c$ in the second case;
it was chosen because it gives $\gamma$ values that are quantitatively close
to those in the first case.  In total, we have taken four cases: two
with $\gamma \propto W/N_s$ ($\gamma = W/N_s$ and $\gamma = 2W/N_s$)
and two with $\gamma\propto W_c/N_s$ ($\gamma = W_c/N_s$, and $\gamma =
2W_c/N_s$).
%

\begin{figure}[tb]
\includegraphics[width=\columnwidth]{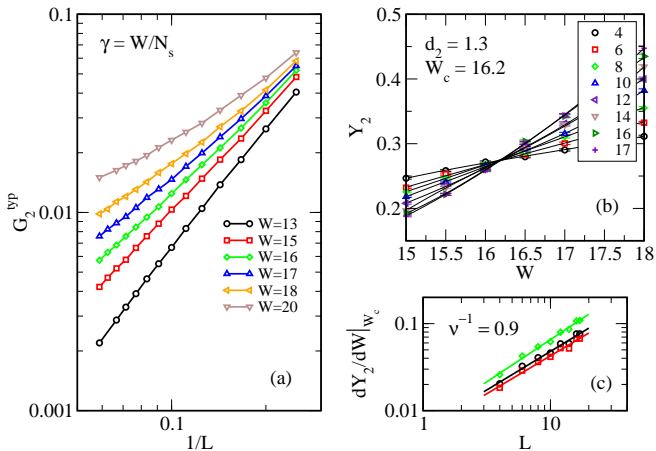}
\caption{(Color online) Scaling of the GIPR for $\gamma=W/N_s$.  (a)
  Plots of $G^\mathrm{typ}_2$ as a function of $L$ for 3000 disorder
  configurations.
  (b) $Y_2$,
  defined by Eq.~(\protect\ref{eq:y2}), for best fit values of $A_0$,
  $d_2$ and $y$.  For these parameters, the critical disorder at which all
  curves cross is $W_c = 16.2$, corresponding to the optimized fitting
  parameters $A_0 = 0.64$, $y=1.6$, and $d_2 = 1.3$.  (c) Plot of
  $dY_2/dW$ at $W=W_c$ (symbols), along with power law fits (solid
  lines) to the data.  The exponents in the fitted curves give
  $\nu^{-1}$, from Eq.~(\ref{eq:dy2}).  The two outlying curves are
  for the extremal values $(d_2,W_c,y)=(1.5,15.9,3.2)$ and $(1.2,16.4,1.2)$ and
  are used to determine uncertainties for $\nu$. The middle curve is
  for the optimized parameters, from which we obtain
  $\nu=1.1$.   Results are summarized in Table~\protect\ref{table:I}.}
\label{fig:fitG1}
\end{figure}

Figure \ref{fig:fitG1}(a) shows the dependence of
$G_2^\mathrm{typ}(0)$ on $L$ for different strengths of disorder
for the case $\gamma = W/N_s$.
  At
short length scales, all the systems are in the critical region 
(albeit in a region where finite size corrections are significant),
and therefore all show similar size dependence.
At long length scales, however, the lines diverge.  For $W<W_c$, the
slope becomes steeper with increasing $L$, consistent with a crossover
to $L^d$ with $d=3$.  For $W>W_c$, the slope decreases with increasing
$L$, consistent with a crossover to a constant value.
This figure suggests that the GIPR is indeed able to distinguish localized
and extended states, even for the relatively small systems studied
here.


We  show that the GIPR displays the same critical
behaviour as the IPR near $W_c$, namely that
\begin{equation}
G_2^\mathrm{typ} = L^{-d_2} \left ( F\left[(W-W_c)L^{1/\nu}\right] + \frac{A_0}{L^y} 
+ \frac{A_1}{L^{2y}} +\ldots \right )
\label{eq:fsscaling}
\end{equation}
where $A_j$ are finite size corrections and $y$ is the critical
exponent for the leading-order irrelevant variable.\cite{Slevin1999}
In all cases, we are able to obtain good scaling behaviour for $4 \leq L \leq
17$ with $A_j=0$ for $j\geq 1$.
%
We thus have five fitting parameters:
$d_2$, $W_c$, $\nu$, $A_0$, and $y$.

\begin{figure}[tb]
\includegraphics[width=\columnwidth]{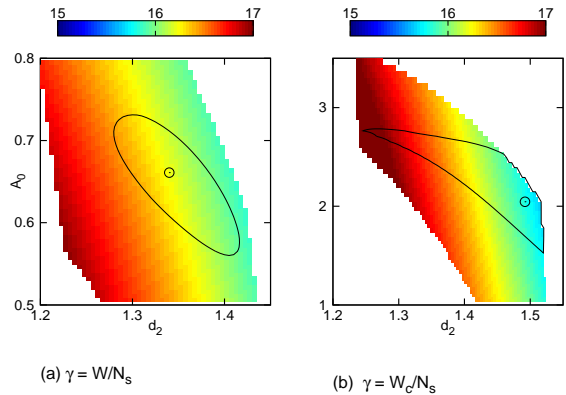}
\caption{(Color online) Critical parameters for (a) $\gamma=W/N_s$ and
  (b) $\gamma=W_c/N_s$ at the best-fit values (a) $y=1.6$ and (b)
  $y=2.9$.  Intensity scale shows the value of $W_X$ that minimizes
  $\chi^2$ locally for each $d_2$ and $A_0$.  Circles indicate
  best-fit values of $d_2$ and $A_0$, obtained from the global minimum
  of $\chi^2$. Black contours bound the regions
  $\chi^2_\mathrm{red}<1$.}  
\label{fig:fitG2}
\end{figure}

We now describe the fitting procedure, using the case $\gamma = W/N_s$
as an example.  Figure \ref{fig:fitG1}(b) shows a plot of
\begin{equation}
Y_2 \equiv G_2^\mathrm{typ} L^{d_2} - \frac{A_0}{L^y}
\label{eq:y2}
\end{equation}
versus $W$ for the optimal values of $A_0$, $y$, and $d_2$. 
 Error bars on
the data are the root-mean-square uncertainty in $G_2^\mathrm{typ}$
due to the finite width of the GIPR distributions (shown, e.g., in
Fig.~\ref{fig:change_gamma}). 
 The solid curves in  Fig.~\ref{fig:fitG1}(b) are
cubic
 fits to the data points. Each pair of curves crosses at a
 different disorder strength, denoted $W_j \pm \delta W_j$, where
 $j\in [1,N_\mathrm{cross}]$ and $N_\mathrm{cross}$ is the the number
 of such crossing points.  (For the 8 curves shown in
 Fig.~\ref{fig:fitG1}(b), there are $N_\mathrm{cross}=28$ crossing
 points.)  The uncertainties $\delta W_j$ are calculated from the
 uncertainties in the fitting parameters.  If the scaling form
 Eq.~(\ref{eq:fsscaling}) holds and the critical parameters are
 correctly chosen, all curves will cross at a single point, $W_X$.
For each $A_0$, $y$ and $d_2$, we find $W_X(A_0,d_2,y)$
by minimizing
\begin{equation}
\chi^2 = \sum_{j=1}^{N_\mathrm{cross}} \left (\frac{W_j - W_X}{\delta W_j}\right )^2.
\label{eq:chi2}
\end{equation}
Plots of $W_X(A_0,d_2,y)$ are shown in Fig.~\ref{fig:fitG2} for optimal values
of $y$ for $\gamma = W/N_s$ and for $\gamma = W_c/N_s$.  We extract
our own  best-fit values for $W_c$ 
from the global minima of $\chi^2(A_0,d_2,y)$, and these are shown as circles
in Fig.~\ref{fig:fitG2}.  A qualitative sense of the goodness-of-fit
can be obtained from Fig.~\ref{fig:fitG1}(b), which is
based on the best-fit parameters for $\gamma=W/N_s$.  
A quantitative measure of goodness-of-fit can be obtained from the
reduced chi-square $\chi^2_\mathrm{red} \equiv \chi^2/(N_c-1)$.
Figure~\ref{fig:fitG2} shows contours around the region of parameter space
 $\chi^2_\mathrm{red} <1$.  In this region, all $Y_2(W)$ curves cross, within
error, at a common point.

The best-fit values
for $W_c$ and $d_2$ are summarized in Table~\ref{table:I}, along with
previously published values.  Quantities in brackets are extremal parameter
values satisfying 
$\chi^2_\mathrm{red}<1$, and are used to estimate the uncertainty in
the critical parameters.  The values for $W_c$ and $d_2$ found from this
analysis are generally within uncertainty of the previously published results.

\begin{table}
\begin{tabular}{c|c|c|c|c}
\hline
\hline
 $\gamma$ & $d_2$ & $W_c$ & $\nu$ & $y$ \\
\hline
$W/N_s$ &1.3 (1.2,1.5) & 16.2 (15.9,16.4) & 1.1 (1.0,1.1) & (1.2,3.2) \\
$2W/N_s$ &1.3 (1.0,1.4) & 16.6 (15.9,17.2) & 1.0 (0.9,1.4) & (1.4,2.5)\\
$W_c/N_s$ &1.5 (1.2,1.5) & 15.8 (15.7,17.0) & 1.3 (1.3,1.4) & (2.1,3.0)\\
$2W_c/N_s$ &1.1 (1.0,1.4) & 17.2 (16.2,17.4) & 1.7 (1.2,1.8) & (1.7,2.2)\\
published &1.3 & 16.54 & 1.57 \\
\hline
\end{tabular}
\caption{Critical parameters from finite-size scaling.  For
  comparsion, previously published results from
  Ref.~\protect\onlinecite{Slevin1999} and
  Ref.~\protect\onlinecite{Mildenberger2002} are shown.  In the first
  column, $W_c$ refers to the accepted value of $16.5$. Numbers in
  parenthesis are estimated bounds on parameters, and are based on the
  parameter regions $\chi^2_\mathrm{red} \leq 1$.}
\label{table:I}
\end{table}

 The next step is to obtain the critical exponent
$\nu$, which is done by fitting a power law to
\begin{equation}
\left. \frac{dY_2}{dW}\right|_{W=W_c} = L^{1/\nu} F^\prime(0).
\label{eq:dy2}
\end{equation}
In  Fig.~\ref{fig:fitG1}(c), we show $dY_2/dW$ at
$W=W_c$, along with power law fits to the data.  The three curves
correspond to the best-fit, minimal, and maximal values of $W_c$ and
$d_2$ shown in Table~\ref{table:I}.  The fitted exponents give three
values of $\nu$ for each $\gamma$, and are shown in the fourth column
of Table~\ref{table:I}. 
We note that $\nu$ is systematically underestimated for $\gamma
\propto W/N_s$, but is closer to the correct answer for $\gamma
\propto W_c/N_s$.  One of the conclusions of this work is that
$\nu$ is more sensitive to the $W$-dependence of $\gamma$ than either
$W_c$ or $d_2$.  This follows directly from the derivative with respect to
$W$ in Eq.~(\ref{eq:dy2}), and means that obtaining an accurate value for
$\nu$ depends on establishing an accurate relationship between $\Delta$ and $W$.


In summary, we have shown that the GIPR can distinguish between
localized and extended states, and moreover that it is possible to
extract critical parameters from a scaling analysis of the GIPR.  The
main issue which arises is how the broadening $\gamma$ influences the
results.  In Table~\ref{table:I}, comparing $\gamma=2W/N_s$ with
$\gamma=W/N_s$ and comparing $\gamma=2W_c/N_s$ with $\gamma=W_c/N_s$,
we see that the results for larger $\gamma$ generally have larger
uncertainties, but do not appear to be systematically shifted towards
or away from their true values.  It thus seems likely that one could
obtain accurate values of $W_c$ and $d_2$ for larger values of
$\gamma$ provided one can study systems that are large enough to keep
the uncertainties to a reasonable size. As mentioned above, one has
the additional requirement that $\gamma$ and $\Delta$ both have the
same dependence on $W$ in order obtain accurate values for $\nu$. This
can be achieved, for example, by taking $\gamma \propto
1/\rho_0(W) N_s$, where $\rho_0(W)$ is the ensemble-averaged density
of states calculated for each strength of disorder.

\section{Conclusions}
\label{sec:conclusions}
We have tested the usefulness of a generalized inverse participation
ratio as a measure of Anderson localization by benchmarking it against
the well-studied case of a disordered three-dimensional tight binding
lattice.  Because the generalized
inverse participation ratio depends on the local density of states,
and not the single particle wavefunctions, it is potentially useful
for studying interacting systems where single particle wavefunctions
are not defined.  We have found that it is possible to extract critical
parameters for the Anderson MIT, and have shown that finite size
effects are not an impediment if the spectral broadening $\gamma$ used to
calculate the local density of states is of the
same order as the level spacing $\Delta$.  

\section*{Acknowledgments}
We acknowledge the support of NSERC of Canada.  This work was made
possible by the facilities of the Shared Hierarchical Academic
Research Computing Network (SHARCNET).


\end{document}